\documentclass[reprint,superscriptaddress,prl,aps,showpacs]{revtex4-1}
\usepackage{dcolumn}
\usepackage{bm}
\usepackage{amsmath}
\usepackage{amsfonts}
\usepackage{amssymb}
\usepackage{bm}
\usepackage[dvips]{graphicx}
\usepackage{fullpage}

\usepackage{subfigure}

\begin{document}
\title{Condensation and Intermittency in an Open Boundary Aggregation-Fragmentation Model}
\author{Himani Sachdeva}
\author{Mustansir Barma}
\affiliation{Department of Theoretical Physics, Tata Institute of Fundamental Research,\\ Homi Bhabha Road, Mumbai-400005, India}
\author{Madan Rao}
\affiliation{Raman Research Institute, C.V. Raman Avenue, Bangalore 560080, India}
\affiliation{National Centre for Biological Sciences (TIFR), Bellary Road, Bangalore 560065, India}

\begin{abstract}
We study real space condensation in aggregation-fragmentation models where the total mass is not conserved, as in phenomena like cloud formation and intracellular trafficking. 
We study the scaling properties of the system with influx and outflux of mass at the boundaries using numerical simulations, supplemented by analytical results in the absence of fragmentation. 
The system is found to undergo a phase transition to an unusual condensate phase, characterized by strong intermittency and
giant fluctuations of the total mass.
A related phase transition also occurs for biased movement of large masses, but with some crucial differences which we highlight.
\end{abstract}
\maketitle

\paragraph*{}
Condensation transitions constitute an important class of non-equilibrium phase transitions, and occur generically in many mass transport
 models \cite{majumdar_rev}  such as the zero range process and its variants \cite{evans_rev}, and the aggregation-chipping model \cite{majumdar}.
  These systems are characterized by a fixed total mass (number of particles) and stochastic rules for exchange of mass between sites.
 When the total mass of the system exceeds a critical value, condensation sets in, with a finite fraction of the total mass forming a macroscopic cluster
that occupies a single site. The phenomenon is akin to Bose condensation, but in real space.

\paragraph*{}
Does the condensation transition survive in a system when the total mass is not conserved, but can  
undergo large fluctuations due to the exit of clusters of all sizes? This question is important in a number of physical situations, ranging from 
formation of clouds and aerosols,  to intracellular trafficking and organelle formation in living 
cells \cite{smoke_etc, golgi_rev, sachdeva}. We address this within a simple but generic 1D model  with aggregation and fragmentation
(chipping) of masses in the bulk, and influx and outflux of masses at the boundaries. Our main finding is that the open system \emph{does} undergo a condensation
transition upon increasing the influx or decreasing the chipping rate. However, the nature of the condensate is very different from that in the closed model \cite{majumdar},
 in that the mass in the condensate shows giant number fluctuations and has a broad distribution, in contrast to the sharply peaked distribution in the closed system \cite{majumdar, rajesh1}.
The condensate, however has a well-defined, finite mean mass for a fixed system size and is thus quite different from the indefinitely growing aggregates in open models which
 allow only single particles to exit at the boundaries \cite{levine, sachdeva}.  
 
 \paragraph*{}
 The intermittent and fluctuating nature of the condensate gives rise to novel signatures: the total mass $M$ itself
 shows giant fluctuations and has a distribution characterized by a prominent non-Gaussian condensate tail whose width scales with system size.
Further, the exit of the condensate from the boundaries and the accompanying sharp drops in $M$ give rise to interesting `charge and fire' behaviour of $M$:
the time series $M(t)$ departs strongly from self-similarity and shows quantitative features of intermittency, which we characterize in terms of
appropriately defined structure functions, as in turbulence phenomena. Turbulence, in the sense of multi-scaling of $n$-point mass-mass correlation
functions has been studied earlier in aggregation models \cite{connaughton}, but our characterization of turbulence-like behaviour is
 quite different, being associated with \emph{temporal} fluctuations of \emph{total} mass. Our results are based on both analytical and numerical work. In the limit of zero chipping,
 we analytically calculate the moments of total mass in steady state, and also the dynamical structure functions, whereas for non-zero chipping, we perform numerical 
 Monte Carlo simulations.

\paragraph*{}
Recently, it has been demonstrated that giant number fluctuations are related to anomalous, non-Porod behaviour of spatial correlation functions in a wide class of systems \cite{dey}.
Our work points to a connection between giant number fluctuations and anomalous dynamical behaviour, namely temporal intermittency, which
is related to higher order correlation functions \emph{in time} \cite{frisch}. It also raises the interesting general question of whether temporal intermittency is present in other systems with 
giant number fluctuations and suggests that dynamical structure functions, as used in the paper, provide a useful probe of intermittency in these systems also.

\paragraph*{}
We work with a general lattice model incorporating diffusion, aggregation, fragmentation, influx and outflux, which goes beyond earlier studies of aggregation with input 
\cite{takayasu, cheng, derrida}, and aggregation and fragmentation in a closed system \cite{majumdar, krapivsky}.
Starting with an empty lattice of $L$ sites at $t=0$, a site $i$ is chosen at random, and one of the following moves occurs: 
\begin{figure} [h]
\centering
{\includegraphics[width=0.48\textwidth]{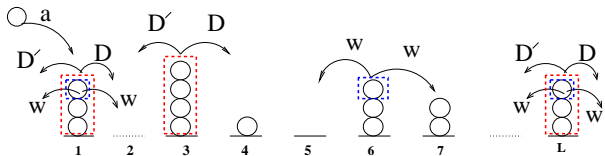}
\label{model}}
\caption {Model: Influx of unit masses at site $1$. Forward and backward stack movement at rates $D$ and $D'$ respectively.
Forward and backward chipping at equal rates $w$. Outflux of full stack or unit mass (via chipping)
from site $L$ (and $1$).}
\end{figure}
\renewcommand{\theenumi}{\roman{enumi}}
\begin{enumerate}
\item Influx: A single particle of unit mass is injected at rate $a$ at the first site ($i=1$).
\item Diffusion and aggregation: With rate $D$ (or $D'$), the full stack on site $i$ (i.e. all particles on the site collectively) hops to
 site $i+1$ (or $i-1$) and adds to the mass already there.
\item Chipping of unit mass: With rate $2w$, a unit mass breaks off from the mass at $i$
and hops to site $i-1$ or $i+1$ with equal probability, adding to the mass already there.
\item Outflux of mass from boundaries: With rate $D$ (or $D'$), the entire mass at site $L$ (or site $1$) exits the system;  with rate
 $w$, a unit mass breaks off from site $L$ (or site $1$) and exits. 
\end{enumerate}

\paragraph*{}
We find that the results depend strongly on two factors: one, whether motion of particles is biased or not and 
two, whether or not exit of masses is allowed from the boundary where influx occurs. In this paper, we
only consider the effect of bias \cite{footnote2}. We find that the occurrence of
the phase transition is robust with respect to bias in the movement of stacks, but not chipping. 
As in the closed periodic case, if the forward and backward chipping rates are unequal, an aggregate is not expected to form \cite{rajesh2}.
Thus, chipping is taken to be unbiased in both the cases
we study in this paper:\\
(A) \emph{Unbiased Stack Hopping}: $D=D'$; exit allowed from sites $1$ and $L$.\\
(B) \emph{Biased Stack Hopping}: $D'=0$; exit allowed from site $L$\\
Influx and chipping occur at rates $a$ and $2w$ respectively in both the above.
\paragraph{\rm{(A)} Unbiased Stack Hopping ($D'=D$):} We discuss both the phases and the critical point below:
\paragraph*{Normal (large $w$) phase:}
In this phase, a typical configuration does not show very large fluctuations about the average mass profile. The total mass $M$ too has normal fluctuations i.e. the rms fluctuations $\Delta M \equiv \sqrt{\langle M^{2}\rangle-\langle M \rangle^{2}} \propto \sqrt{L}$, with the distribution
for the rescaled mass variable $(M-\langle M\rangle)/\Delta M$, approaching a Gaussian at large $L$. The mass 
distribution $P(m,j,L)$ at a given site $j$ is found to depend on $j$ and $L$ only through the rescaled position variable $x=j/L$ \cite{supplement}, implying that for a given $x$, 
all moments of mass are independent of $L$ to leading order.
\paragraph*{Condensate (small $w$) phase:} A typical configuration deviates strongly from the average profile, with the largest (local) fluctuations scaling as system size $L$.
On monitoring the largest mass $m_{1}$ in the system, we find that its average value $\langle m_{1} \rangle$ is proportional to $L$ \cite{supplement}, 
 implying that the system contains a macroscopic condensate. The presence of the condensate has a strong effect on all steady state properties of the system such as the total mass $M$, mass at
a site, etc. The probability distributions of all these quantities have an exponential tail with a characteristic
 mass $M_{0}$ where  $M_{0}\propto L$ for a given $w$ and $a$.  We refer to this exponential tail 
 as the `condensate' tail and describe below, how it appears in various steady state distributions:
\begin{figure} [t]
\centering
{\includegraphics[width=0.4\textwidth]{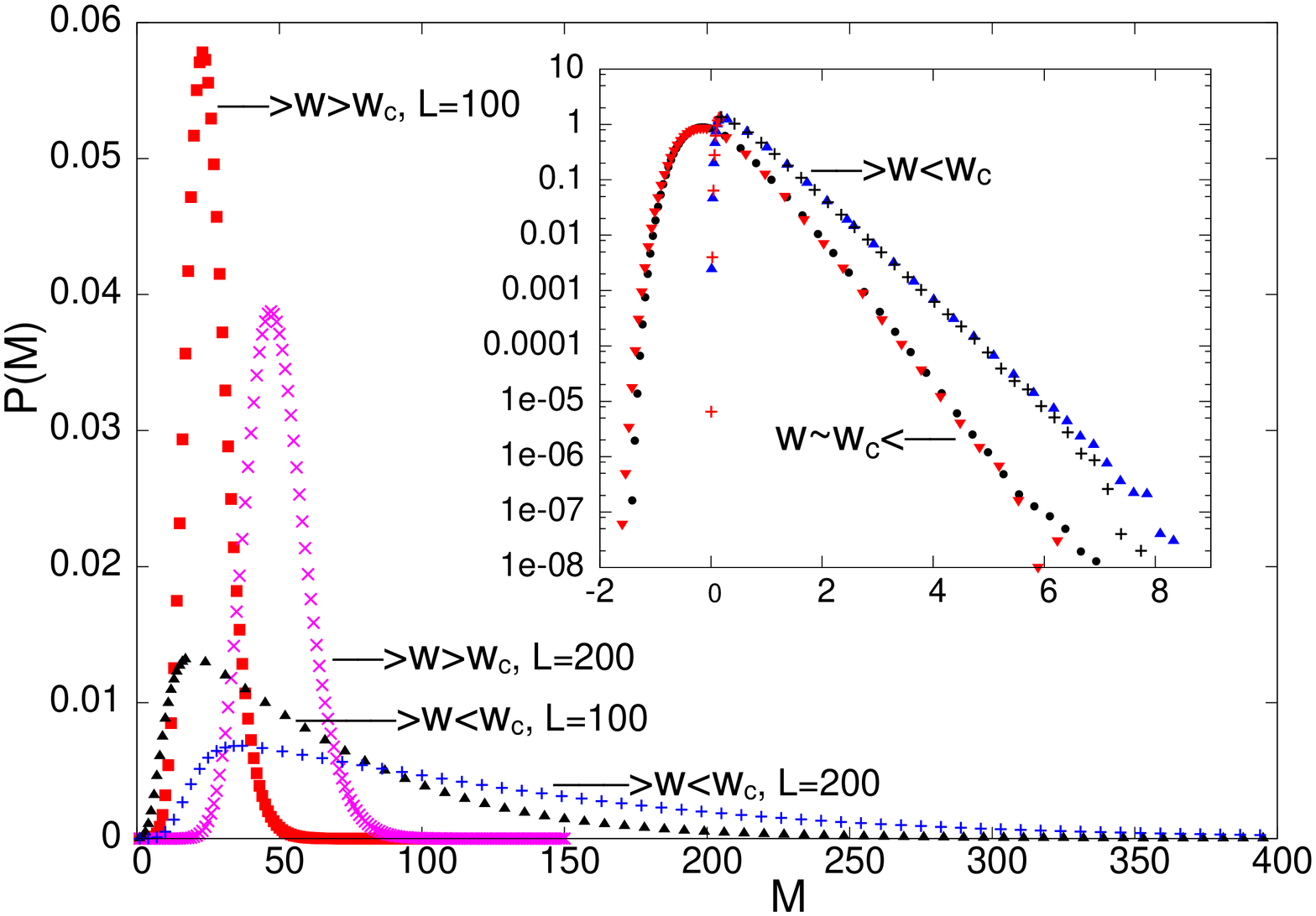}}
\caption{ $P(M)$ vs. $M$ for $L=100$ and $L=200$ in the normal phase ($a=1$, $D=0.75$, $w=2$) and condensate phase ($a=1$, $D=0.75$, $w=0.25$). Inset: Scaling collapse of tails
on plotting $LP(M)$ vs. $M/L$ in the condensate phase and $L^{2/3}P(M)$ vs. $(M-\langle M\rangle)/L^{2/3}$ 
near the critical point ($a=1$, $D=0.75$, $w=1.5$).}
\label{pdist1}
\end{figure}
\renewcommand{\theenumi}{\roman{enumi}}
\begin{enumerate}
 \item The steady state distribution $P(M)$ of total mass
 of $M$ in the system behaves as $P(M) \sim \frac{1}{M_{0}}\exp \left(-\frac{M}{M_{0}}\right)$ at large $M$  
[fig. \ref{pdist1}] . Consequently, the rms fluctuation $\Delta M$ of total mass shows non-Gaussian 
behaviour, scaling as $L$ rather than $\sqrt{L}$. We have analytically calculated various moments of the total mass in the limit $w=0$ \cite{supplement}.
We find that $\Delta M /L\simeq 0.46(a/D)$, in the limit of large $L$, which agrees 
well with numerics.
\item The distribution $P(m_{1})$ of the largest mass $m_{1}$ also follows $P(m_{1}) \sim \frac{1}{M_{0}}\exp \left(-\frac{m_{1}}{M_{0}}\right)$ for large $m_{1}$ \cite{supplement}.
\item The distribution of masses exiting from the left or right boundary \cite{supplement} is found to follow: 
$P_{exit}(m) \sim \frac{1}{L^{2}}\left(\frac{1}{M_{0}}\exp\left(-\frac{m}{M_{0}}\right)\right)$, 
for large $m$ \cite{footnote3}. 
\item The single site mass distribution $P(m,j,L)$  \cite{supplement} follows 
$\frac{1}{L}f\left(\frac{j}{L}\right)\left(\frac{1}{M_{0}}\exp\left(-\frac{m}{M_{0}}\right)
\right)$ at large $m$ \cite{footnote3}.
 The factor $1/L$ arises as the aggregate can be at any one of the $L$ sites, 
and $f(j/L)$ reflects that the aggregate does not visit all sites with the same probability.
The rms fluctuation $\Delta m(x,L)$ of mass at a given $x=j/L$ is thus anomalously large as well:
it increases as $\sqrt{L}$ with $L$ rather than being $\ensuremath{\mathcal{O}\!(1)}$, as in the normal
 phase. 
\end{enumerate}

\paragraph*{}
 That there is no constraint on the total number of particles per site in our model is crucial for $L$-dependent fluctuations to arise.
Systems such as vibrated needles \cite{narayan} and passive particles in fluctuating fields \cite{das, mishra} also display giant number fluctuations, 
but in these systems, fluctuations in a region of linear size $\Delta l$ depend primarily on $\Delta l$, rather than $L$ \cite{dey}.
This is traceable to hard core interactions between particles in these models. Once this constraint
 is removed, macroscopic stacks can form and mass fluctuations depend on $L$ \cite{nagar}.

\paragraph*{Critical point $w_{c}$:} The transition from the normal to the condensate phase takes 
place at a critical chipping rate $w_{c}$, which increases with injection rate $a$ if $D$ is held constant \cite{footnote4}. 
At criticality also, large fluctuations of the total mass are found, consistent with
 $\Delta M \propto L^{\theta_{c}}$ where $\theta_{c}\simeq2/3$. The mass distribution has a tail of the form: 
$P(M) \sim \frac{1}{M_{2}}\exp\left(\frac{-(M-\langle M\rangle)}{M_{2}}\right)$ where $\langle M\rangle\sim L$ and $M_{2}\sim 
 L^{\theta_c}$ [inset, fig. \ref{pdist1}] . Interestingly, we find that there is a similar $L$-dependent tail in the distribution of masses exiting 
from the left, but not the right.
 This is presumably because although an $L$-dependent aggregate forms close
 to the left boundary, it dissipates due to chipping on diffusing through the bulk of the system, and 
does not survive up to the right boundary. 

\paragraph*{}
Contrasting signatures of the phases also appear in dynamical properties: $M(t)$ is self-similar in
time in the normal phase but exhibits breakdown of self-similarity in the condensate phase. The 
breakdown of self-similarity is captured in the behaviour of the structure
 functions $S_{n}(t)=\langle \left[ M(t)-M(0)\right]^{n} \rangle$ \cite{footnote5} where $\langle ...\rangle$ denotes
 average over histories. Self-similar signals typically show $S_{n}(t) \propto t^{\gamma n}$ as $t/\tau\rightarrow 0$, where
 $\gamma$ is a constant and $\tau$ is a time scale which characterizes the lifetime of the largest structures in the system.
 A deviation from $S_{n}(t) \propto t^{\gamma n}$ reflects the breakdown
 of self-similarity and may occur, 
for example, if the signal $M(t)$ alternates between periods of quiescence (small or no activity) and bursts (sudden large changes) \cite{frisch}.
Such an alternation is characteristic of intermittency.
 The most well-studied measures of intermittency are the flatness,
 defined as $\kappa (t)= S_{4}(t)/S_{2}^{2}(t)$; and the hyperflatness $h(t)= S_{6}(t)/S_{2}^{3}(t)$. For intermittent
signals, both  $\kappa(t)$ and $h(t)$ diverge as $t/\tau\rightarrow 0$ \cite{frisch}. Below we present evidence for intermittency
in our model.
\begin{figure} [h]
\centering
\subfigure[]{
\includegraphics[width=0.22\textwidth]{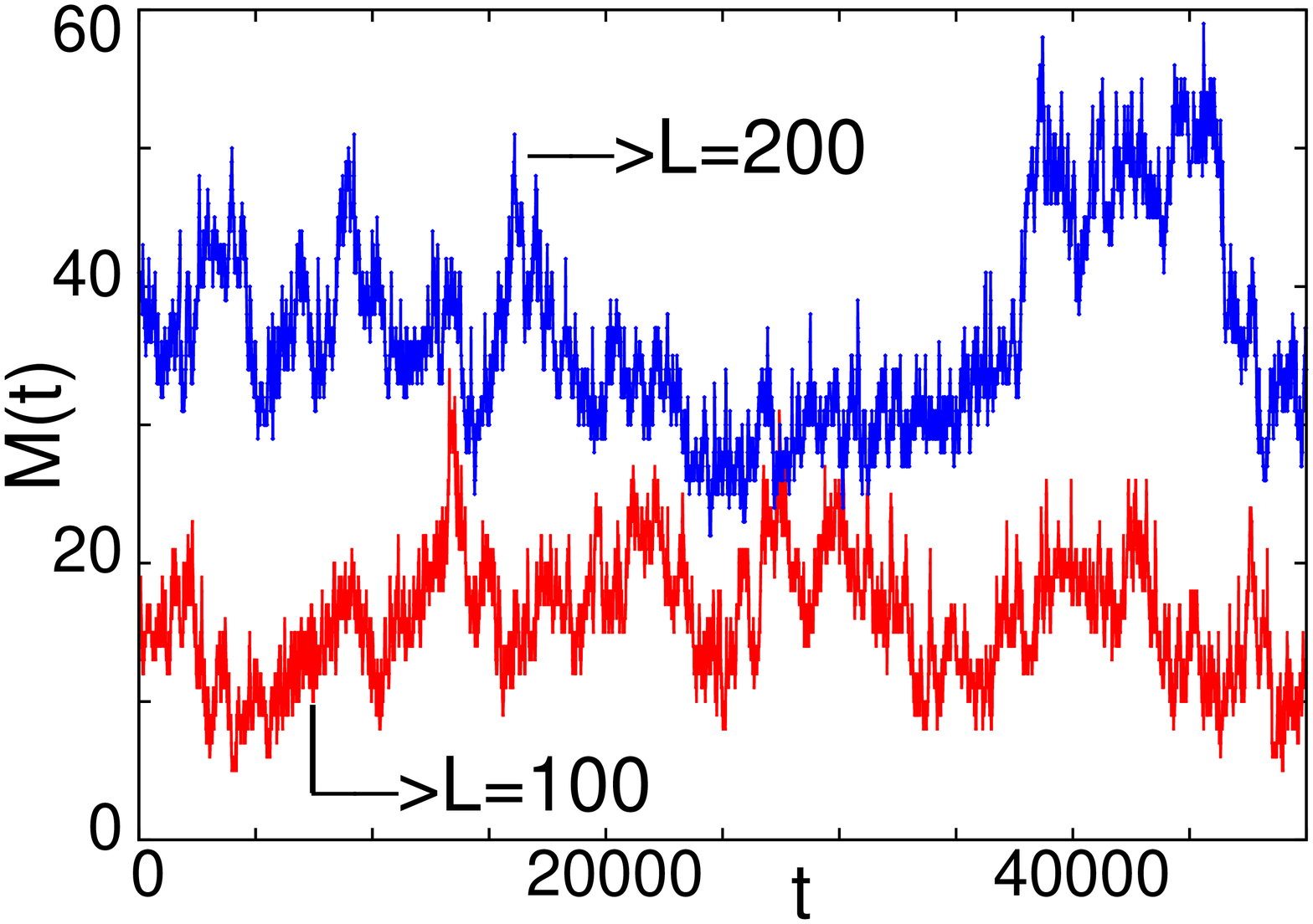}
\label{tseries2}}
\subfigure[]{
\includegraphics[width=0.22\textwidth]{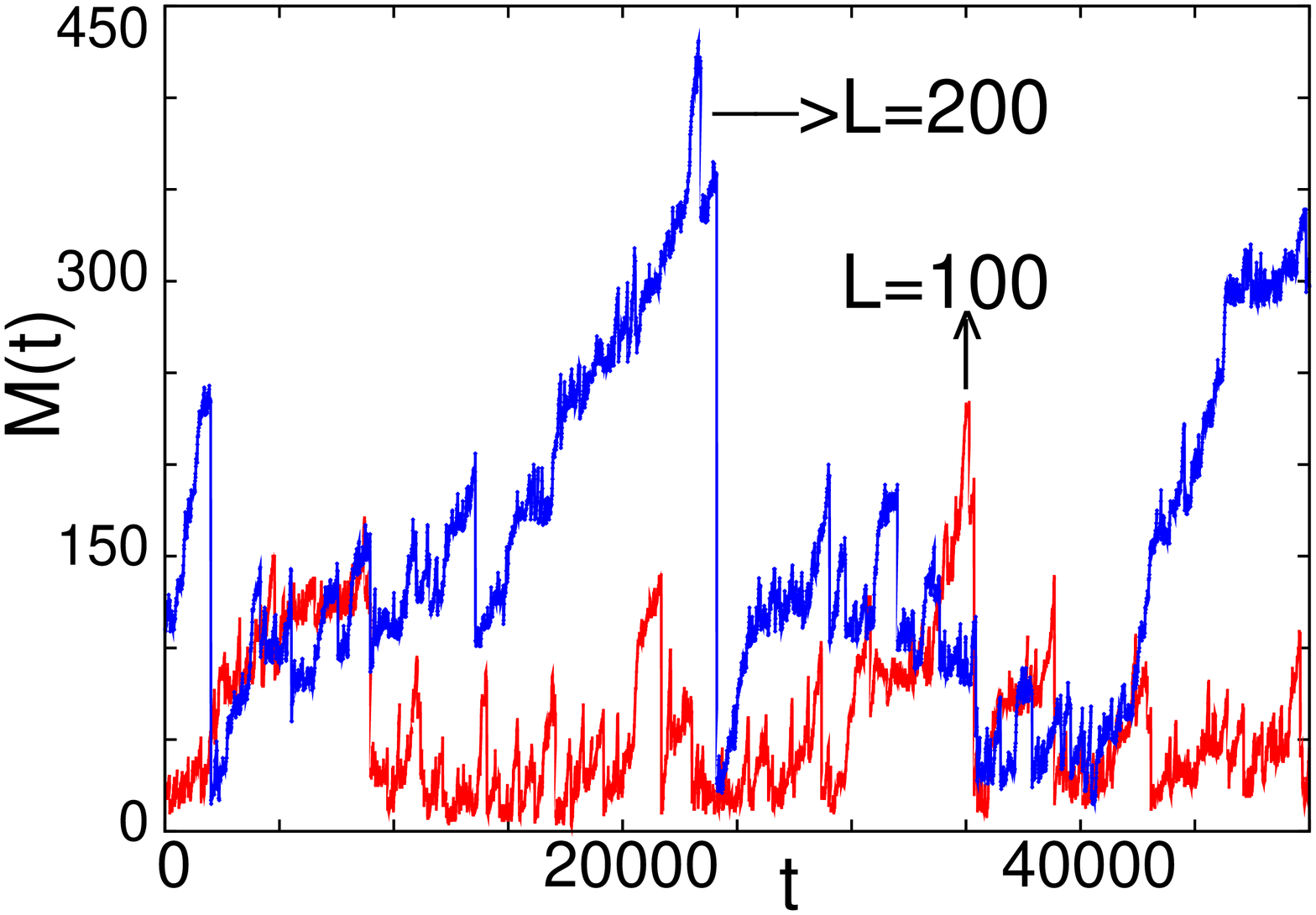}
\label{tseries1}}
\subfigure[]{
\includegraphics[width=0.44\textwidth]{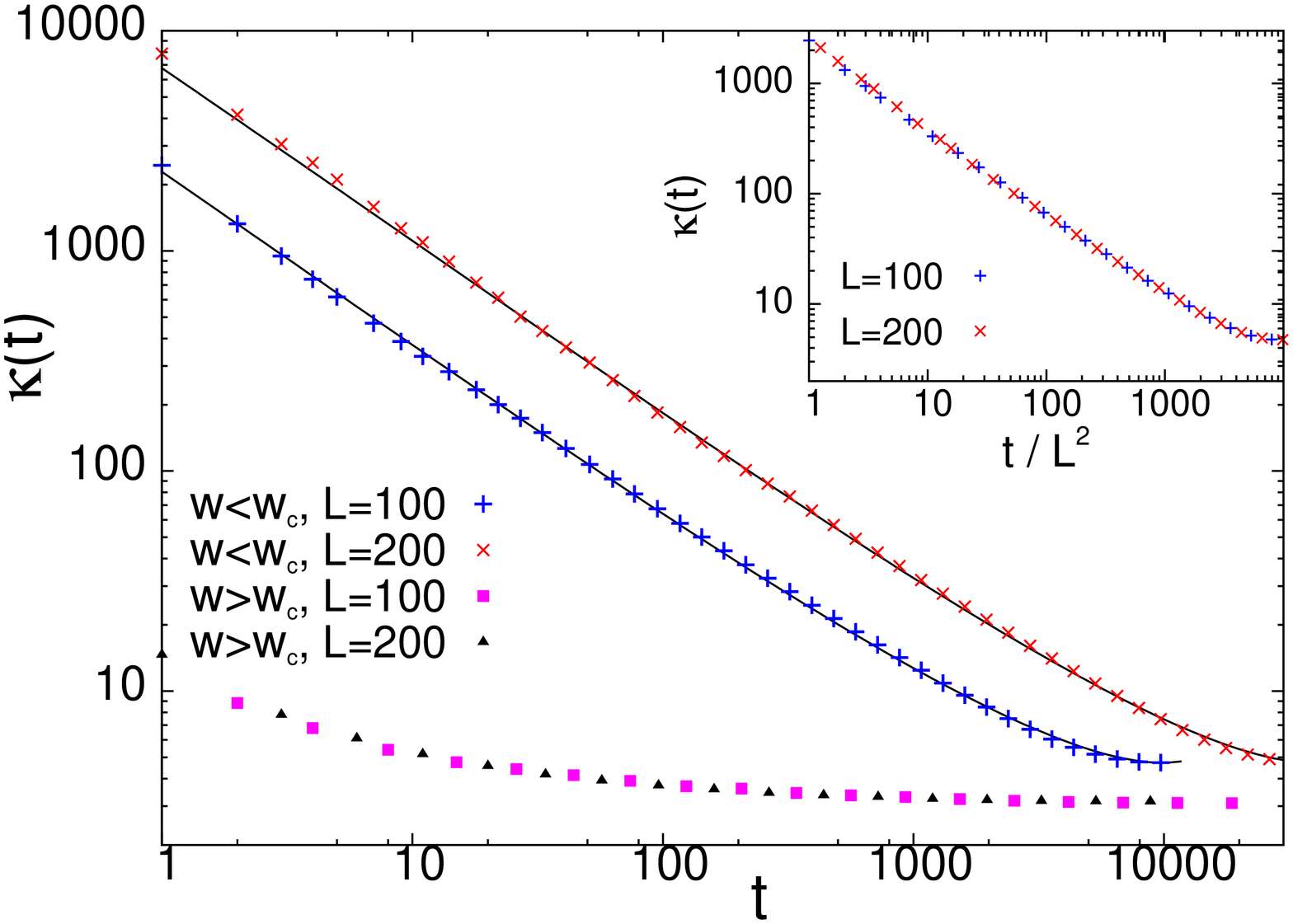}
\label{flatness}}
\caption{(a)-(b): Realizations of $M(t)$ vs. $t$ for different $L$ in the (a) normal phase ($a=1$, $D=0.75$, $w=3.0$)
and (b) condensate phase ($a=1$, $D=0.75$, $w=0.25$). Note that the y-axis in (a) and (b) has a different scale. \newline (c) $\kappa(t)$ vs. $t$ for $L=100$
and $L=200$ in the two phases. Solid lines are fits to the form described in the text for $t\ll L^{2}$ for $w<w_{c}$. Inset: Scaling collapse of $\kappa(t)$ vs. $t$ for different $L$
on scaling time as $t/L^{2}$ in the condensate phase.}     
\label{tseries}
\end{figure}
\paragraph*{Normal Phase:}
In this phase, the structure functions are independent of $L$ at small $t$ and scale as 
$S_{2n} \sim t^{\beta_{n}}$ where the dependence of $\beta_{n}$ on $n$ is close to linear \cite{supplement}, indicating 
self-similarity of the time series $M(t)$ [fig. \ref{tseries2}]. The flatness $\kappa (t)$ and hyperflatness
$h(t)$ approach a finite, $L$ independent value as $t\rightarrow 0$ [fig. \ref{flatness} and fig. 4 in \cite{supplement}].
\paragraph*{Condensate Phase:}
In the condensate phase, $M(t)$ builds up as mass is injected and drops as masses exit, with occasional 
large crashes [fig. \ref{tseries1}] corresponding to the exit of condensates with $\ensuremath{\mathcal{O}\!(L)}$ mass.
The structure functions are found to scale as: $S_{n} \sim L^{n} f_{n}(t/L^{2})$, where $f_{n}$ is consistent with the form $f_{n}(y)\sim (-1)^{n}yg_{n}[\log(y)]$  for small $y$, (with $g_{n}$
chosen to be a polynomial) \cite{footnote6}, 
 and approaches an $n$-dependent constant value at large $y$ \cite{supplement}. 
Thus, the system shows strong intermittency: at small $t$, all structure functions $S_{n}$ behave as $\sim t$ with the $n$-dependence 
entering only through the multiplicative $\log t$ terms. 
It follows that $\kappa(t)$ and $h(t)$ diverge at small times in a strongly $L$ dependent way [fig. \ref{flatness} and fig. 4 in \cite{supplement}].
In fact, they are functions of $t/L^{2}$ and diverge as $t/L^{2}\rightarrow 0$ [inset, fig. \ref{flatness} and inset, fig. 4 in \cite{supplement}]. 
We have also analytically calculated $S_{2}(t)$ in the zero chipping limit $w=0$ \cite{supplement} and find that it agrees well with numerical results.

\paragraph*{Critical point $w_{c}$:}  $M(t)$ continues to show intermittency at the critical point with flatness and hyperflatness diverging
 as $t\rightarrow 0$ in an $L$ dependent manner. However, there seems to be no simple scaling which collapses the curves for different $L$.
\paragraph*{}
\paragraph{\rm{(B)}. Biased stack hopping ($D'=0$):}
The steady state can be obtained exactly in the limiting cases of only chipping $D=0$ \cite{levine} and only aggregation $w=0$ \cite{jain}.
The probability distribution of the rescaled mass, $(M-\langle M\rangle)/\Delta M$, is Gaussian
 with $\Delta M \propto \sqrt{L}$, in both limits but for different reasons. 
In the pure chipping limit $D=0$, this follows from the independence of masses at different sites, implied by the product
 measure of the mass distribution \cite{levine}; in the pure stack hopping limit
$w=0$, on the other hand, it is associated with the formation and exit of aggregates of typical size $\sqrt{L}$ \cite{jain, reuveni}. 
This essential difference is well captured by the time series data. For $w=0$, the total mass $M$ shows intermittency on time scales of order
 $\sqrt{L}$, corresponding to a typical time interval of ${\mathcal{O}\!(\sqrt{L})}$ between exit events. Flatness and hyperflatness are functions of $t/\sqrt{L}$ and diverge as power laws as
 $t/\sqrt{L}\rightarrow 0$. By contrast, for $D=0$, the time series $M(t)$ is not
 intermittent. Thus, intermittency rather than anomalous steady state fluctuations of $M$,
 is a key signature of aggregate formation when stack hopping is driven.
\begin{figure} [h]
\centering
{\includegraphics[width=0.48\textwidth]{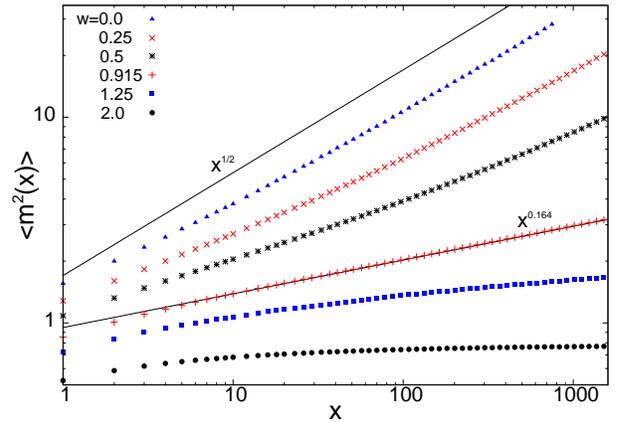}}
\caption {$\langle m^{2}(x) \rangle$ vs. $x$ for $a=1$, $D=1.5$ and different $w$. Note the upward (downward)
bending of curves on log-log plot in aggregation (chipping) dominated phase. There is no bending at $w_{c}$.}
\label{mxdd}
\end{figure}

\paragraph*{}
As $w$ is decreased, there is a phase transition from a normal phase to an aggregation-dominated phase characterized by intermittency. Unlike the unbiased case, however,
the typical size of aggregates that exit the system is now expected to scale as $\sqrt{L}$ rather than $L$. This is consistent with the behaviour of $\langle m^{2}(x) \rangle$ vs. $x$ 
[fig. \ref{mxdd}].
For large $w$, the plots of $\langle m^{2}(x)\rangle$ approach a constant value at large $x$, thus indicating that there are no $x$ dependent aggregates
at large $x$ and no intermittency. For small $w$, the plots of $\langle m^{2}(x) \rangle$ vs. $x$ bend upwards,
 consistent with an approach to $\sqrt{x}$ at large $x$. Exit of $\sqrt{L}$ sized aggregates leads to intermittency \cite{supplement}, 
as for $w=0$. The transition takes place at $w_{c}$, (corresponding to the curve with no bending on the log-log plot), at which $\langle m^{2}(x)\rangle$
behaves as $\sim x^{\alpha_{c}}$ with $\alpha_{c}\simeq0.16$. $M(t)$ shows intermittency at the critical point also.
 
\paragraph*{}
In conclusion, the principal result of this work is to establish the existence of a condensate phase in unbiased aggregation-chipping
 models where total mass is not conserved due to influx and outflux at the boundaries. This phase is characterized by anomalous steady state fluctuations of the total mass, 
and by intermittency in the dynamics, as quantified by the divergence of the flatness. It is likely that flatness would be a  useful measure in other mass exchange models also.
The phase transition also occurs when the movement of stacks is biased, but the intermittent, aggregation-dominated phase in this case is different.
 
\paragraph*{Acknowledgements:} We thank D. Dhar for useful comments on the manuscript.


\begin{thebibliography}{99}
\bibitem{footnote3}  This holds for most $w<w_{c}$. For $w$ very close to $w_{c}$, no condensate tail is visible in 
$P_{exit}(m)$ for right-exiting masses and in $P(m,j,L)$ for $j$ in a small region near the right boundary.
\bibitem{footnote5} This is in analogy with structure functions of the velocity field in fluid flow (see e.g. \cite{frisch})
\end{thebibliography}
\end{document}


\title{Condensation and Intermittency in an Open Boundary Aggregation-Fragmentation Model-- Supplementary Material}

\begin{abstract}
In this supplement, we provide the following details of analytical and numerical results for the model with unbiased stack hopping: (i) an outline of the calculation of $\langle M^{n}\rangle$ and $S_{2}(t)$ in the limit $w=0$ of this model
(ii) further numerical evidence for the differences between the condensate and normal phases, as described in the main paper
(iii) a theoretical estimate of the critical chipping rate $w_{c}$. We also provide numerical evidence for intermittency in the aggregation-dominated phase of the model with biased
stack hopping.
\end{abstract}

\maketitle
\section{Unbiased stack movement}
\subsection{A. Some analytical results in the limit $w=0$:}
\paragraph*{}
The condensate phase extends in the range $0\leq w<w_{c}$. In the zero chipping limit $w=0$, where stack movement is the only dynamical move in the bulk, we are able to obtain some analytical 
results for the moments of total mass in the steady state, and the structure function $S_{2}(t)$. These calculations are outlined below \cite{footnote1}.
\paragraph*{Calculation of moments $\langle M^{n}\rangle$ in steady state:} To calculate moments of total mass $M$, we consider mass between sites $i+1$ and $j$, i.e. $m_{i,j}=\sum\limits_{k=i+1}^{j} m_{k}$ (where $m_{k}$ is the mass on the $k^{th}$
site) and the corresponding probability distribution $P_{i,j}(m)$ of this mass. In steady state, $P_{i,j}(m)$ satisfies the following equation:
\begin{subequations}
\begin{equation}
\begin{split}
& P_{i,j+1}(m)+P_{i,j-1}(m)+P_{i-1,j}(m)+P_{i+1,j}(m)\\
& -4P_{i,j}(m)=0 \qquad 0<i, \quad i+1<j, \quad j<L
\end{split}
\end{equation}
\begin{equation}
\begin{split}
& a[1-\delta_{m,0}]P_{0,j}(m-1)-aP_{0,j}(m)+D[P_{0,j+1}(m)\\
&+P_{0,j-1}(m)+P_{1,j}(m)-3P_{0,j}(m)]=0 \\
& \qquad \qquad \qquad \qquad \quad 1<j, \quad j<L
\end{split}
\end{equation}
\begin{equation}
\begin{split}
P_{i,L-1}(m)+P_{i-1,L}(m&)+P_{i+1,L}(m)- 3P_{i,L}(m)=0\\
&0<i, \quad i+1<L
\end{split}
\end{equation}
\begin{equation}
\begin{split}
2\delta_{m,0}+P_{i,i+2}(m)+&P_{i-1,i+1}(m)-4P_{i,i+1}(m)=0\\
& 0<i, \quad i+1=j<L
\end{split}
\end{equation} 
\label{Pij}
\end{subequations}
To obtain $\langle M^{n}\rangle$, we now do the following: \\(i) obtain recurrence relations satisfied by the generating function $Q_{i,j}(z)=\sum\limits_{m=1}^{\infty} P_{i,j}(m)z^{m}$ from the above 
equation (ii) differentiate $Q_{i,j}(z)$ $n$ times to get recurrence relations for the $n^{th}$ mass moments $\langle m_{i,j}^{n}\rangle$ (iii) convert the recurrence relations satisfied
by $\langle m_{i,j}^{n}\rangle$ to a differential equation satisfied by $\langle m_{xy}^{n}\rangle$ by taking the continuum limit $x=i/L$, $y=j/L$ (iv) solve this differential equation (which
turns out to be a Laplace equation on a triangular region) to get $\langle m_{xy}^{n}\rangle$ (v) substitute $x=0$, $y=1$ in $\langle m_{xy}^{n}\rangle$ to get
$\langle M^{n}\rangle$. This gives the following expressions for $\langle M^{n}\rangle$, to leading order in $L$:

\begin{equation}
\begin{split}
&\langle M\rangle =0.5\frac{aL}{D}, \qquad \langle M^{2}\rangle\sim 0.461\left(\frac{aL}{D}\right)^{2},\\
& \Delta M=\sqrt{\langle [M-\langle M\rangle]^{2}\rangle} \sim 0.459\frac{aL}{D}\\
&\langle M^{3}\rangle\sim 0.615\left(\frac{aL}{D}\right)^{3}, \qquad \langle M^{4}\rangle\sim 1.074\left(\frac{aL}{D}\right)^{4}
\end{split}
\end{equation}

\paragraph*{Calculation of structure function $S_{2}(t)$:}
\begin{equation}
\begin{split}
S_{2}(t)&=\langle [M(t)-M(0)]^{2}\rangle\\
&=2[\langle M^{2}\rangle-\langle M \rangle^{2}]-2[\langle M(0)M(t)\rangle-\langle M \rangle^{2}]
\end{split}
\end{equation}
We have already calculated $\langle M^{2} \rangle$
and $\langle M \rangle$. To obtain the auto-correlation function $\langle M(0)M(t)\rangle-\langle M \rangle^{2}$, we calculate the quantity $G_{i,j}(t)=\langle M(0)m_{i,j}(t)\rangle-
\langle M \rangle\langle m_{i,j} \rangle$. This satisfies the equation:
\begin{subequations}
\begin{equation}
\begin{split}
\frac{\partial{G_{i,j}(t)}}{\partial{t}}=&D[\langle G_{i+1,j}(t)+G_{i-1,j}(t)+G_{i,j+1}(t)\\
&+G_{i,j-1}(t)-4G_{i,j}(t) \rangle]
\end{split}
\end{equation}
\begin{equation}
\begin{split}
&G_{0,j}(t)=G_{1,j}(t) \qquad G_{i,L+1}(t)=G_{i,L}(t) \\
&\qquad G_{i,i}(t)=0 
\end{split}
\end{equation}
\begin{equation}
\begin{split}
G_{i,j}(0)&=\langle Mm_{i,j}\rangle-\langle M \rangle\langle m_{i,j} \rangle\\
&=\frac{1}{2}\left[\langle m_{0,j}^{2}\rangle -\langle m_{0,i}^{2}\rangle+
\langle m_{i,L}^{2}\rangle-\langle m_{j,L}^{2}\rangle\right]\\
&\quad-\langle M \rangle\langle m_{i,j} \rangle
\end{split}
\end{equation}
\end{subequations}
As before, we take the continuum limit $x=i/L$, $y=j/L$ in space to get a differential equation which we solve to obtain $G_{xy}(t)$ and hence $S_{2}(t)$. This gives the following expression for
$S_{2}(t)$:
\begin{equation}
\begin{split}
 S_{2}(t)=\sum\limits_{n=1,3,5...}^{\infty} &\left\{\frac{16\eta^{2}}{(n \pi)^{4}}\left(n\pi\coth\left[\frac{n\pi}{2}\right]-1\right) 
+\frac{8\eta}{(n\pi)^{2}}\right\}\\
&\times\left\{1- \exp\left[-\frac{D\pi^{2} n^{2}t}{L^{2}}\right]\right\}
\end{split}
\label{S2t}
\end{equation}
To extract the small $t$ behaviour, we convert the above sum into an integral and take the limit $\tau=\frac{Dt}{L^{2}}\rightarrow 0$. In this limit, the $\coth$ term in the above sum 
behaves asymptotically as $-\tau\log[\tau]$. Thus, we obtain the following small $t$ expression for $S_{2}(t)$
\begin{equation}
 S_{2}(t)\sim -A_{0}t\log\left(A_{1}\frac{Dt}{L^{2}}\right)
 \label{S2smallt}
\end{equation}
 where $A_{0}$ and $A_{1}$ are some constants. Higher order structure functions can also be calculated similarly, but the calculations become very cumbersome.

\subsection{B. Numerical evidence}
\paragraph*{}
The condensate and normal phases in the model are characterized respectively by the presence or absence of a macroscopic condensate with typical mass which is 
$\ensuremath{\mathcal{O}\!(L)}$ for a system of $L$ sites. Signatures of this condensate appear in both the steady state and dynamical properties of the system. Below, we briefly
recapitulate these properties, as presented in the main paper, and provide additional numerical evidence for them.\newline
\subsubsection{Steady state} In the condensate phase, the following steady state mass distributions have an exponential tail with characteristic mass $M_{0}\propto L$,
which we refer to as the condensate tail:
\begin{figure} [h]
\centering
{\includegraphics[width=0.45\textwidth]{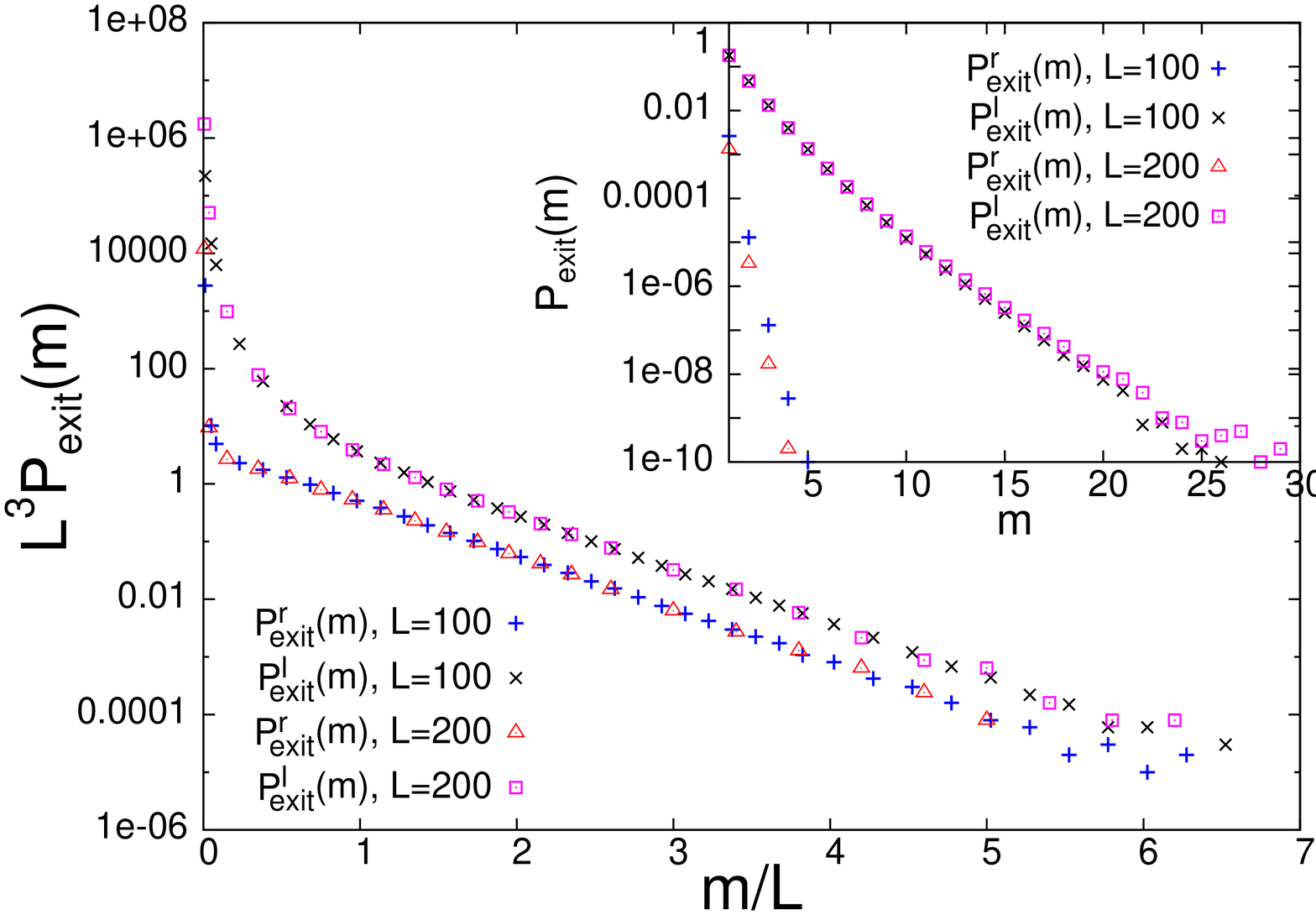}}
\caption{Exit currents:  In the condensate phase ($a=1$, $D=0.75$, $w=0.25$), $P^{l/r}_{exit}(m)$ show condensate tails with good scaling collapse to $L^{3} P^{l/r}_{exit}(m)$ vs. $m/L$ for different $L$.
Inset: In the normal phase ($a=1$, $D=0.75$, $w=3.0$), exit currents do not have condensate tails.}
\label{pexit}
\end{figure}
\begin{figure} [h]
\centering
{\includegraphics[width=0.45\textwidth]{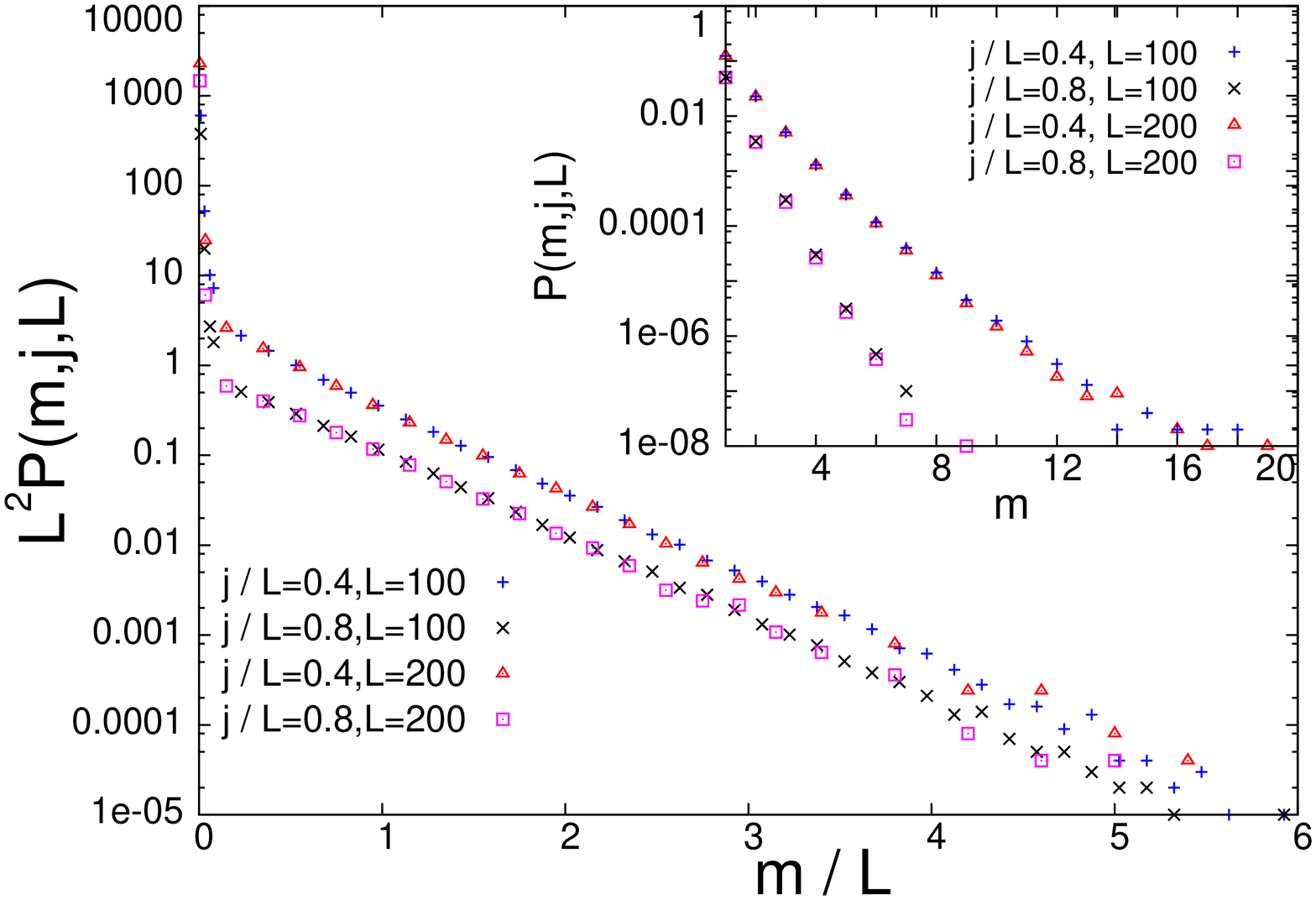}}
\caption{Single site mass distributions: In the condensate phase ($a=1$, $D=0.75$, $w=0.25$), $P(m,j,L)$ has condensate tails with the same 
characteristic $M_{0}$ for different $j/L$. There is good scaling collapse of $L^{2} P(m,j,L)$ vs. $m/L$ for a given $j/L$. Inset: In the normal phase ($a=1$, $D=0.75$, $w=3.0$), $P(m,j,L)$ for a given $j/L$
is independent of $L$ to leading order and shows no condensate tail.}     
\label{pj2}
\end{figure}
\begin{figure} [h]
\centering
{\includegraphics[width=0.45\textwidth]{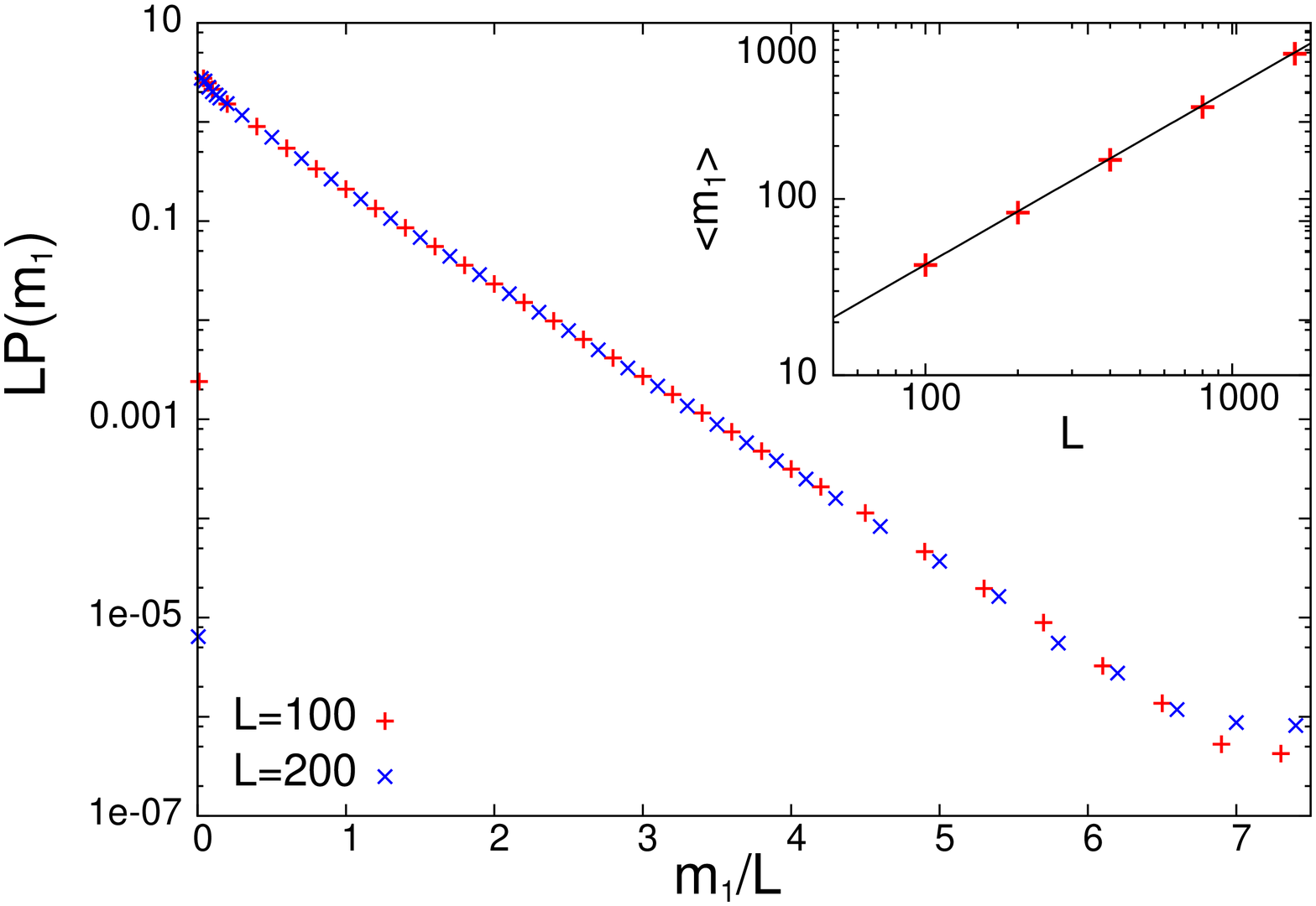}}
\caption{Probability distribution of largest mass $m_{1}$ in the condensate phase ($a=1$, $D=0.75$, $w=0.25$): $P(m_{1})$ has condensate tails with good scaling collapse to 
$LP(m_{1})$ vs. $m_{1}/L$. Inset: $\langle m_{1}\rangle$ scales linearly with $L$.}     
\label{largestm}
\end{figure}

\renewcommand{\theenumi}{\roman{enumi}}
\begin{enumerate}
 \item $P(M)$: The probability distribution $P(M)$ of the total mass $M$ in the system obeys $P(M) \sim \frac{1}{M_{0}}\exp \left(-\frac{M}{M_{0}}\right)$ 
 for large $M$ where $M_{0}\propto L$. Evidence for this is shown in fig. 2 of main paper.
 \item $P_{exit}(m)$: The distributions $P^{l}_{exit}(m)$ and $P^{r}_{exit}(m)$ of masses $m$ exiting from the left and the right boundaries (i.e. sites $1$ and $L$)
 respectively obey: $P^{l/r}_{exit}(m) \sim \frac{A_{l/r}}{L^{2}}\left(\frac{1}{M_{0}}\exp\left(-\frac{m}{M_{0}}\right)\right)$ at large $m$ where the prefactor $A_{l/r}$ is different 
 for the two distributions. Figure \ref{pexit} shows the scaling collapse of the tails of the two distributions on plotting $L^{3}P^{l/r}_{exit}(m)$ vs. $M/L$.
 \item $P(m,j,L)$: The mass distribution at the $j^{th}$ site of the system obeys 
 $P(m,j,L)\sim\frac{1}{L}f\left(\frac{j}{L}\right)\left(\frac{1}{M_{0}}\exp\left(-\frac{m}{M_{0}}\right)\right)$ at large $m$.
 Figure \ref{pj2} shows the scaling collapse on plotting $L^{2}P(m,j/L=0.4)$ and $L^{2}P(m,j/L=0.8)$ vs. $m/L$. As implied by the form of $P(m,j,L)$, for large $m$, the single site distributions
 for different $j/L$ collapse on to different curves, which differ from each other only by the constant prefactor $f(j/L)$.
 \item $P(m_{1})$: The mass distribution of the largest mass $m_{1}$ behaves as $P(m_{1}) \sim \frac{1}{M_{0}}\exp \left(-\frac{m_{1}}{M_{0}}\right)$ for large $m_{1}$ [fig. \ref{largestm}].
  As a result, the largest mass in the system is macroscopic (on an average), scaling as $\langle m_{1}\rangle \propto L$ [inset, fig. \ref{largestm}]
\end{enumerate}

By contrast, in the normal phase, we have:
\begin{enumerate}
 \item $P(M)$: The distribution for the rescaled mass variable $(M-\langle M\rangle)/\Delta M$, approaches a Gaussian at large $L$ (see fig. 2 of main paper).
 \item $P_{exit}(m)$: To leading order in $L$, $P^{l}_{exit}(m)$ is independent of $L$, while $P^{r}_{exit}(m)\sim \left(b/L\right)^{m}$ where $b$ is a constant.
 (See inset in fig. \ref{pexit}). This behaviour of the exit currents can be established analytically for the limiting case $D=0$ \cite{levine}.
 \item $P(m,j,L)$: The mass distribution $P(m,j,L)$ at a given site $j$ depends on $j$ and $L$ only through the rescaled position variable $x=j/L$. The inset in fig. \ref{pj2} shows 
 $P(m,j/L=0.4)$ and $P(m,j/L=0.8)$ vs. $m$ for different $L$. Note that there is no significant dependence on $L$ for a given $j/L$.
 \item $\langle m_{1}\rangle$ is sub-linear in $L$, such that $\langle m_{1}\rangle/L\rightarrow 0$ for large $L$.
 \end{enumerate}
\subsubsection{Dynamical properties}
As discussed in the main paper, a key signature of the condensate phase is intermittency in the time series of the total mass $M$. This is captured by the small $t$ behaviour of
the structure functions $S_{n}(t)$:
\begin{figure} [h]
\centering
\subfigure[]{
\includegraphics[width=0.45\textwidth]{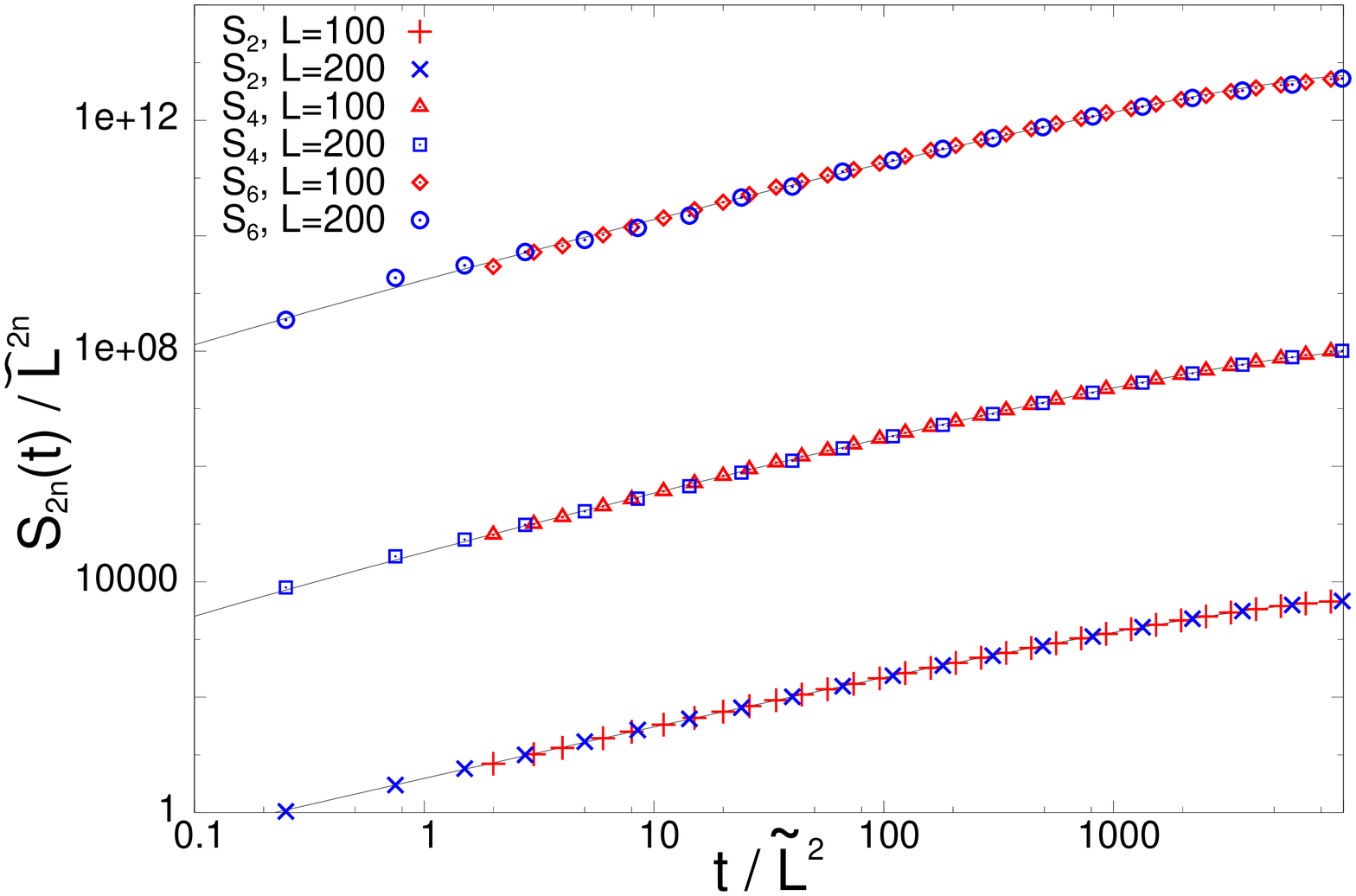}
\label{snagg}}
\subfigure[]{
\includegraphics[width=0.45\textwidth]{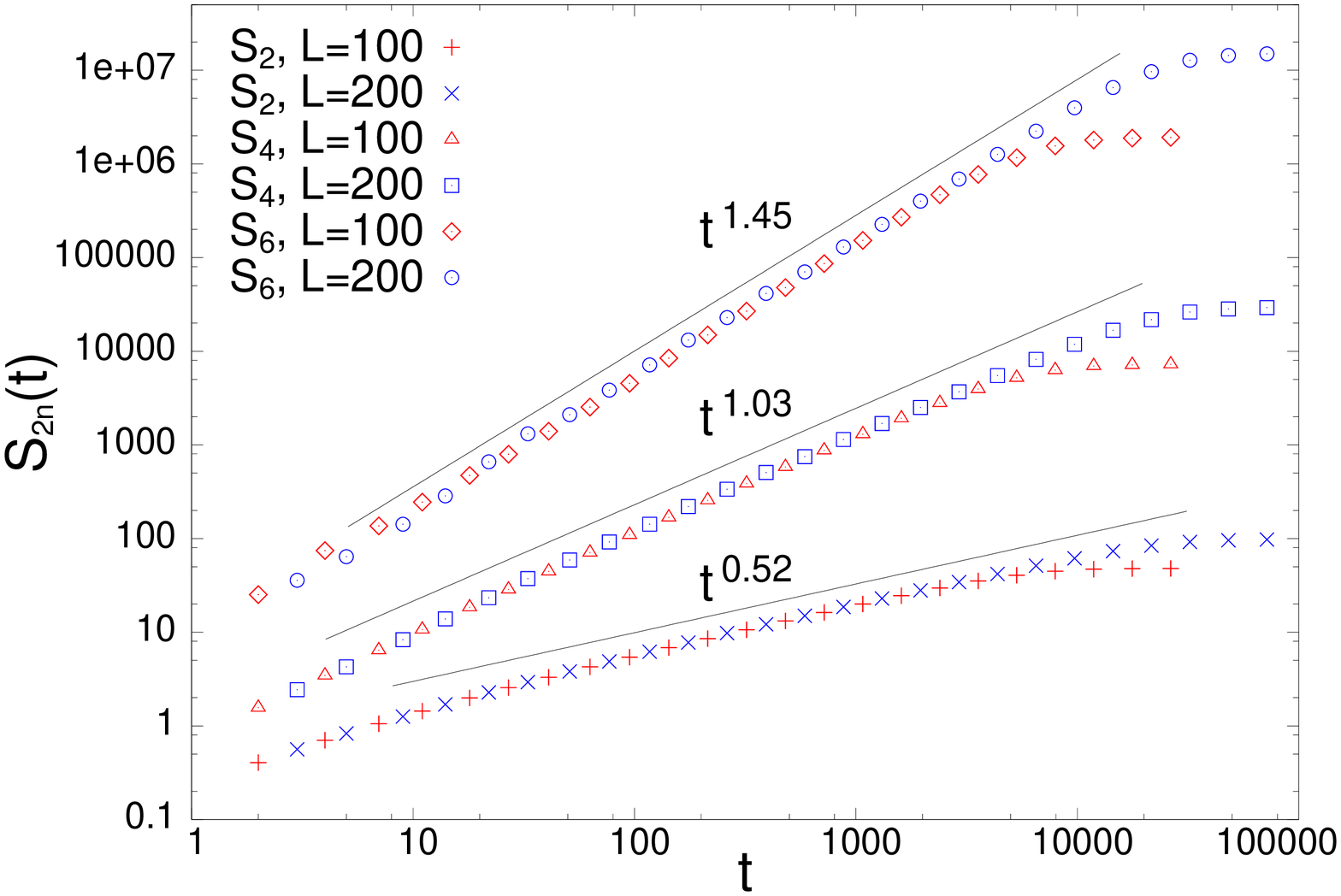}
\label{snnorm}}
\caption{Structure functions: (a) In the condensate phase ($a=1$, $D=0.75$, $w=0.25$), structure functions for different $L$ show good scaling collapse to $S_{2n}(t)/\tilde{L}^{2}$ vs $t/\tilde{L}^{2}$.
Solid lines indicate $tg_{2n}[log(t/L^{2})]$ as described in text.\newline (b) In the normal phase ($a=1$, $D=0.75$, $w=3.0$),  $S_{2n}(t)$ show no dependence on $L$ and behave as
$\sim t^{\beta_{2n}}$ (solid lines) for small $t$.}
\end{figure}
\begin{figure} [h]
\centering
{\includegraphics[width=0.45\textwidth]{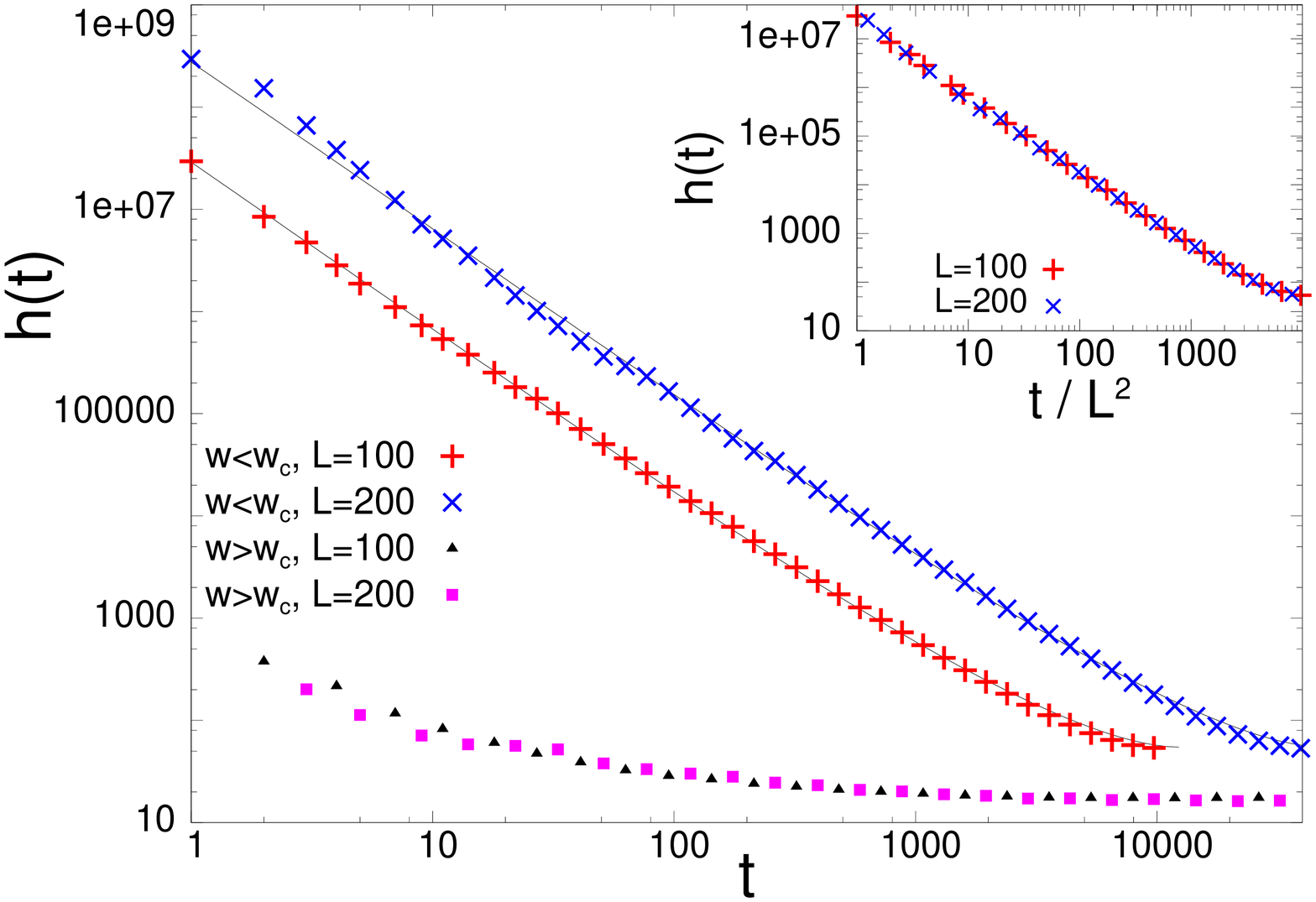}}
\caption{$h(t)$ vs. $t$ for $L=100$ and $L=200$ in the condensate ($a=1$, $D=0.75$, $w=0.25$) and normal ($a=1$, $D=0.75$, $w=3.0$) phases. Solid lines are fits to the form described in the text for $t\ll L^{2}$ for $w<w_{c}$.
Inset: Scaling collapse of $h(t)$ vs. $t$ for different $L$ on scaling time as $t/L^{2}$ in the condensate phase.}
\label{hyper}
\end{figure}

\paragraph*{Condensate phase:} Numerics show that the structure functions scale with $L$ as:
$S_{n} \sim L^{n} f_{n}(t/L^{2})$. Motivated by the small $t$ form for $S_{2}(t)$ in the limit $w=0$ \eqref{S2smallt}, we use the form $f_{n}\sim (-1)^{n}yg_{n}[\log(y)]$  for small $y$, and get good fits 
with $g_{n}$ chosen as polynomials. Figure \ref{snagg} shows the scaling collapse of the $n^{th}$ order structure functions for
two different $L$ to the above form for $n=2,4,6$. We plot $S_{2n}/\tilde{L}^{2}$ vs $t/\tilde{L}^{2}$ (where the rescaling $\tilde{L}=L/100$ has been done
to just display all three structure functions in the same plot clearly). The solid lines are fits to $tg_{2n}[\log(t/L^{2})]$ where $g_{2}$, $g_{4}$ and $g_{6}$
are taken to be polynomials of degrees $1$, $2$ and $3$ respectively. The above forms of $S_{n}$ imply that the flatness $\kappa(t)=S_{4}(t)/S_{2}^{2}(t)$
and hyperflatness $h(t)=S_{6}(t)/S_{2}^{3}(t)$ diverge at small $t$. The divergence of the flatness in the condensate phase is shown in fig. 3(c) of the main paper.
Figure \ref{hyper} shows the divergence of the hyperflatness, with the solid lines obtained from the fitted forms of $S_{6}$ and $S_{2}$. 
The inset in fig. \ref{hyper} shows the scaling collapse of $h(t)$ for different $L$ on scaling
time as $t/L^{2}$.
\paragraph*{Normal phase:}
At small times ($t\ll L^{2}$), the structure functions $S_{n}(t)$ are independent of $L$ and behave as 
$S_{n} \sim t^{\beta_{n}}$ (except very close to $t=0$), with the dependence of $\beta_{n}$ on $n$ being close to linear. Figure \ref{snnorm} shows $S_{2}$,
$S_{4}$ and $S_{6}$ for two different $L$ along with solid lines $t^{\beta_{n}}$ with $\beta_{2}\sim 0.52$, $\beta_{4}\sim 1.03$
and $\beta_{6}\sim 1.45$. The small deviation from $\beta_{n}\propto n$ may be due to sub-leading correction terms.
Thus, in this phase, flatness (see fig. 3(c) of main paper) and hyperflatness (fig. \ref{hyper}) approach finite, $L$ independent values as $t\rightarrow 0$.

\subsection{C. Estimation of critical point $w_{c}$}
\paragraph*{}
Let us define $\rho_{j}$ and $s_{j}$ respectively as the average density and occupation probability (probability that the site is non-empty)
at a site $j$. 
In steady state, the balance of average mass current at each site  gives the following exact expression (with $D$ set equal to $1$):
\begin{equation}
\label{eqn:currenteq}
 \rho_{j} + ws_{j}=a\left(1- j/L\right)
\end{equation}
We obtain a rough analytical estimate for $w_{c}$ as follows: the
 system is in the condensate phase as long as there is at least a finite fraction of the lattice that \emph{locally}
 satisfies conditions for aggregate formation. As in the closed periodic case \cite{majumdar}, we take these conditions to be: (i) the occupation probability of a site
 in this region is equal to a critical occupation probability $s_{c}(w)$
 and (ii) the mass density is greater than a critical mass density $\rho_{c}(w)$. While we do not have the forms of $s_{c}$ and $\rho_{c}$ in the open boundary case,
 in the limit $L\rightarrow\infty$, we approximate these by the exact conditions derived
 for the closed periodic case \cite{rajesh1}:
\begin{subequations}
\begin{equation}
s_{j}=s_{c}(w)=(\sqrt{1+w} -1)/(\sqrt{1+w} +1)
\end{equation}
\begin{equation}
\rho_{j}\geq\rho_{c}(w) \qquad \text{where} \quad \rho_{c}(w)=\sqrt{1+w} -1
\end{equation}
\end{subequations} 
Numerical simulations suggest that this is a reasonable approximation, at least away from the right boundary $x=L$.
At $w_{c}$, a vanishingly small fraction ($j/L\rightarrow0$) of the system satisfies conditions for aggregate formation. 
Thus, setting $s_{j}=s_{c}(w)$, $\rho_{j}=\rho_{c}(w)$ and $j/L \rightarrow 0$ in \eqref{eqn:currenteq}, we get:
\begin{equation}
w_{c}=\left(2a-1+\sqrt{4a+1} \right)/2
\end{equation}
This agrees with the numerically determined $w_{c}$, obtained from $\Delta M$ vs. $L$ curves, to within $10\%$. 
\section{Biased Stack Movement}
\begin{figure} [h]
\centering
{\includegraphics[width=0.48\textwidth]{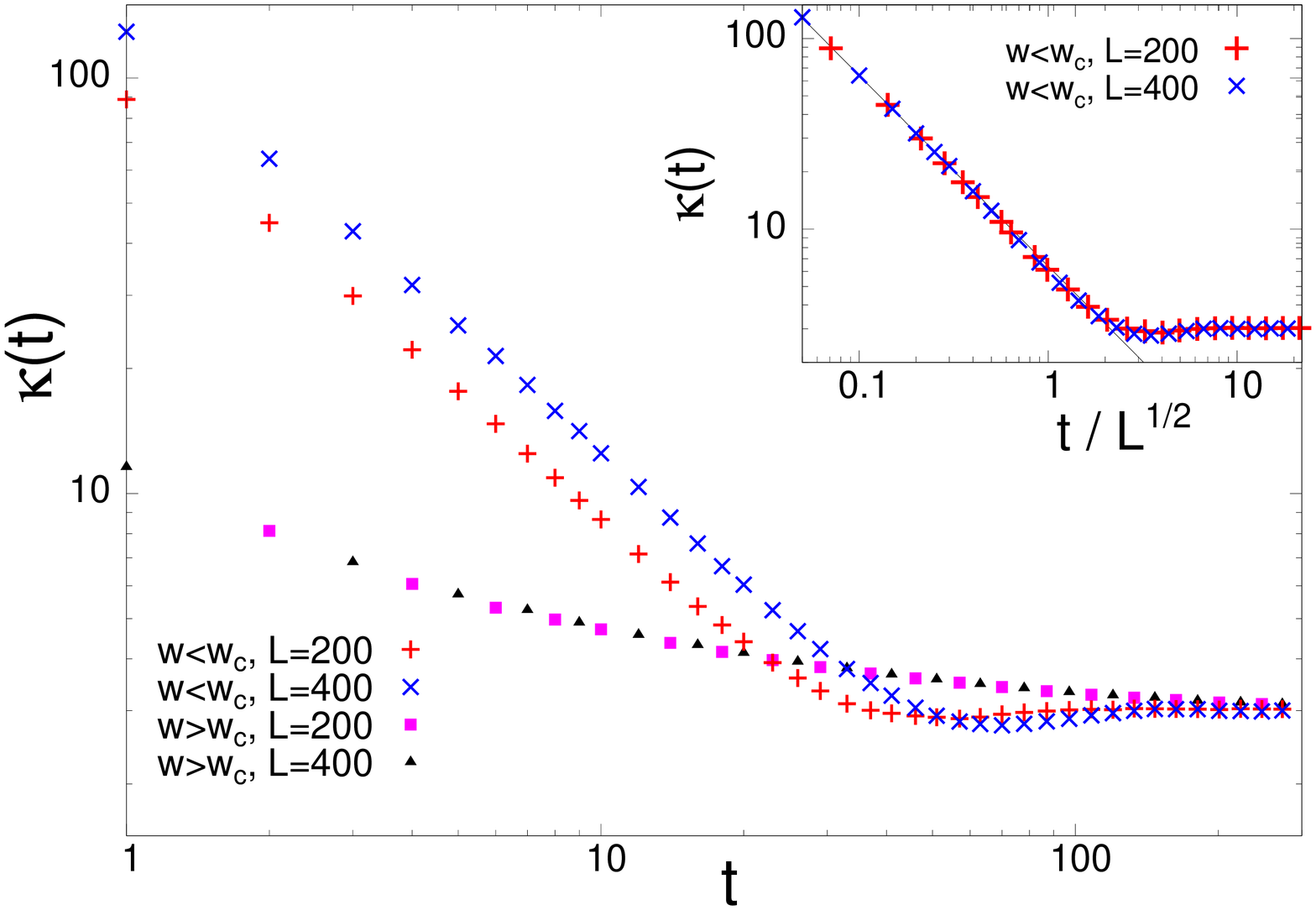}}
\caption {Biased stack Movement:  $\kappa(t)$ vs. $t$ for $L=200$ and $L=400$ in the aggregation-dominated ($a=1$, $D=1.5$, $w=0.125$) and normal ($a=1$, $D=1.5$, $w=3.0$) phases.
Inset: Scaling collapse of $\kappa(t)$ vs. $t$ for different $L$
on scaling time as $t/\sqrt{L}$ in the aggregation-dominated phase. The divergence at small $t$ roughly follows $\sim (t/\sqrt{L})^{-1}$ (solid line).}  
\label{flatdriv}
\end{figure}
When stack movement is biased, the basic signature of the aggregation-dominated phase is the intermittency of $M(t)$, as the steady state mass distribution $P(M)$ 
has the same (Gaussian) form in both phases, as explained in the main paper. Figure \ref{flatdriv} shows the flatness $\kappa(t)$ for two different values of $L$ in both the aggregation-dominated
and normal phases. In the aggregation-dominated phase, the flatness $\kappa(t)$ diverges
as $t\rightarrow0$ in an $L$ dependent manner while it approaches a constant $L$-independent value in the normal phase.
The inset in fig. \ref{flatdriv} shows the scaling collapse of $\kappa(t)$ for different $L$ on rescaling time as $t/\sqrt{L}$. Thus, intermittency occurs on
time scales of order $\sqrt{L}$ when stack hopping is biased.